# Vacancy-tuned paramagnetic/ferromagnetic martensitic transformation in Mn-poor $Mn_{1-x}CoGe$ alloys

**(Short title: Temperature window in $Mn_{1-x}CoGe$ alloys)**

E. K. Liu[1], W. Zhu[1], L. Feng[1], J. L. Chen[1], W. H. Wang[1], G. H. Wu[1(a)], H. Y. Liu[2], F. B. Meng[2], H. Z. Luo[2] and Y. X. Li[2]

[1] *Beijing National Laboratory for Condensed Matter Physics, Institute of Physics, Chinese Academy of Sciences, Beijing 100190, China*

[2] *School of Material Science and Engineering, Hebei University of Technology, Tianjin 300130, China*

**Abstract** - It is shown that a temperature window between the Curie temperatures of martensite and austenite phases around the room temperature can be obtained by a vacancy-tuning strategy in Mn-poor $Mn_{1-x}CoGe$ alloys ($0 \le x \le 0.050$). Based on this, a martensitic transformation from paramagnetic austenite to ferromagnetic martensite with a large magnetization difference can be realized in this window. This gives rise to a magnetic-field-induced martensitic transformation and a large magnetocaloric effect in the $Mn_{1-x}CoGe$ system. The decrease of the transformation temperature and of the thermal hysteresis of the transformation, as well as the stable Curie temperatures of martensite and austenite, are discussed on the basis of the Mn-poor Co-vacancy structure and the corresponding valence-electron concentration.

(a) E-mail: ghwu@aphy.iphy.ac.cn

The magnetic-field-induced martensitic transformation and the magnetocaloric effect in various kinds of systems have increasingly been studied for sensing [1-4] and magnetocaloric applications [5,6]. However, in the equiatomic MM'X (M, M' = transition metals, X = Si, Ge, Sn) compounds, such functional characters have been scarcely reported. MnCoGe, as a typical MM'X compound, possesses magnetic and structural transitions and has been studied for decades [7-10]. Stoichiometric MnCoGe crystallizes in the hexagonal $Ni_2In$-type austenite (*A*) phase and, at about 420 K, in the paramagnetic state, it undergoes a martensitic structural transformation to, also paramagnetic, orthorhombic TiNiSi-type martensite (*M*) [7,8,11]. The austenite and the martensite phases have Curie temperatures $T_C^A$ =273 K and $T_C^M$ =355 K [8,11], respectively, both phases being collinear ferromagnets [9,12]. Experiments and calculations have shown that the *M* phase carries a larger magnetic moment than the *A* phase [8,13]. As NiAs-$Ni_2In$-type compound, hexagonal off-stoichiometric MnCoGe is prone to form transition-metal vacancies at its 2d sites [14]. Recently, it has been found that the Co vacancies decrease the martensitic-transformation temperature ($T_m$) of MnCoGe and, in $Mn_{1.07}Co_{0.92}Ge$, a field-induced martensitic transformation from ferromagnetic austenite to ferromagnetic martensite (FM/FM-type) has been reported [15,16]. In addition, various studies [10,15,17,18] have shown that the Curie temperatures of both martensite and austenite depend intensively on the Co content in Co-poor off-stoichiometric alloys.

According to the aforementioned, one may expect that, if $T_m$ could be appreciably lowered, while both $T_C^A$ and $T_C^M$ would be kept approximately unchanged, a martensitic transformation from paramagnetic austenite to ferromagnetic martensite (PM/FM-type transformation), similar to that in $Fe_2MnGa$ [19,20], may be obtained inside a temperature window limited by $T_C^A$ and $T_C^M$ . This may lead to a larger magnetization difference ($\Delta M$) between the two phases at the field-induced martensitic transformation and, simultaneously, because of equivalent exothermic/endothermic behavior at the magneto-structural transition, to a large magnetocaloric effect. Very recently, Hamer *et al.* [21] have substituted Ge for Sn in MnCoSn and found a fixed temperature window, in which a PM/FM-type martensitic transformation takes place. This has made the $MnCoGe_{1-x}Sn_x$ system potentially interesting as field-induced shape memory alloys and as magnetocaloric material.

In this letter, we report that it is possible to realize a quite similar temperature window around room temperature by tuning the Co-vacancy concentration in Mn-poor $Mn_{1-x}CoGe$ alloys. It will be shown that the transformation temperature $T_m$ can be tuned in the temperature window between $T_C^A$ and $T_C^M$ and that the desired PM/FM-type martensitic transformation is accompanied by a considerable $\Delta M$.

From a high-temperature x-ray diffraction (XRD) study on $Mn_{0.98}CoGe$ [22], it is known that Co atoms partially occupy the Mn-deficient sublattice and that vacancies at Co sublattices are formed instead. In the present study of the $Mn_{1-x}CoGe$ system, we consider the Mn-vacancy and Co-vacancy structures and calculate their total energies in the $Ni_2In$-type parent phase using an *ab initio* method. The results confirm that the Co-vacancy structure $(Mn_{1-x}Co_x)(Co_{1-x}\square_x)Ge$, is energetically favourable. We may thus attempt to take advantage of this Co-vacancy structure by properly reducing the Mn content to tune the phase transition in $Mn_{1-x}CoGe$ alloys. One can see that the Co-vacancy content is equal to the deficiency level ($x$) of Mn in the $Mn_{1-x}CoGe$ alloys.

Polycrystalline ingots of $Mn_{1-x}CoGe$ ($x$=0, 0.010, 0.020, 0.030, 0.035, 0.045 and 0.050) were prepared by arc-melting pure metals in high-purity argon atmosphere. The ingots were annealed at 1123 K for five days and quenched in cool water. The crystal structures of the specimens were identified by powder XRD analysis with Cu-$K_\alpha$ radiation. Magnetization measurements were carried out in a superconducting quantum interference device (SQUID). Also, differential thermal analysis (DTA) with heating and cooling rates of 2.5 K/min was used to investigate the martensitic transformation.

Figure 1 shows the XRD patterns obtained at different temperatures for the alloy with $x$=0.035. At 320 K, the $Ni_2In$-type structure is detected with lattice parameters a=4.0850 and c=5.3145 Å. At 220 K, the Bragg reflections of the TiNiSi-type structure are well indexed with the lattice parameters a=5.9323, b=3.8246 and c=7.0433 Å. The unit-cell volume has increased by 4.0%, which means a huge lattice distortion during the transformation. The crystal structures of both phases are also shown in fig. 1. XRD examinations at room temperature (not shown) confirm that the alloys with the compositions of $x$ = 0 ~ 0.045 all exhibit the martensitic transformation with $T_m$ monotonously decreasing in a temperature region around room temperature. It shows that the value of $T_m$ in $Mn_{1-x}CoGe$ is sensitive enough to the Mn-deficiency level ($x$), i.e. the Co-vacancy concentration, to be tailored to a desired temperature.

Figure 2 shows the *M(T)* curves of $Mn_{1-x}CoGe$ ($x$ = 0.010, 0.030, 0.035 and 0.045) alloys in a low magnetic field of 0.01 T and the characteristic parameters $T_m$, $T_C$ and the thermal hysteresis ($\Delta T$) are listed in table 1. In fig. 2, $T_C^M$ =353 K at the high-temperature side corresponds to the Curie temperature of the *M* phase for the alloy with vacancy content $x$ = 0.010, while $T_C^A$ =261 K at the low-temperature side corresponds to that of the *A* phase for the alloy with vacancy content $x$ = 0.045. These values of $T_C^M$ and $T_C^A$ determine a temperature interval. It can be seen that the martensitic transformations of these two alloys, occurring at 365 K and 242 K (see table 1) respectively, are both outside the temperature interval. Further measurements of the alloys with $x$=0 and 0.020 show that their $T_C^M$ values lie, with slight variation, at the highest border temperature, the $T_C^M$ value for x = 0.010. Similarly, the $T_C^A$ value of the alloy with $x$ = 0.050 also coincides with the lowest border temperature, the $T_C^A$ value for x = 0.045. This clearly shows that the $T_C$ values of martensite and austenite are both insensitive to the Mn-deficiecy level ($x$) in $Mn_{1-x}CoGe$ alloys with $x$ between 0 and 0.050. Thus, in the Mn-poor $Mn_{1-x}CoGe$ system, there exists a temperature window with a width of about 100 K between $T_C^M$ and $T_C^A$ in which martensitic transformations occur. This situation is very similar to that in $MnCoGe_{1-x}Sn_x$ alloys [21]. It implies that there exists an intrinsic temperature window in MnCoGe, which can be uncovered by different tailoring methods. Collecting these results, a phase diagram for the $Mn_{1-x}CoGe$ system ($0 \leq x \leq 0.050$) is proposed in the inset of fig. 2. For vacancy contents in the range $0.010 < x < 0.045$, the alloys exhibit a first-order martensitic transformation from the PM *A* phase to the FM *M* phase. Here, one should note that this coincidence of the structural and the magnetic transitions is quite different from the situation in $Ni_2MnGa$ system, in which $T_m$ merges with $T_C^A$ at the same temperature [23]. The results for two alloys with vacancy contents $x$ = 0.030 and 0.035 are shown in fig. 2. During cooling of these alloys, the paramagnetic *A* phase transforms to the ferromagnetic *M* phase above the hypothetical ordering temperature $T_C^A$ so that the $T_C^A$ has become irrelevant. Likewise, during heating, the ferromagnetic *M* phase transforms to paramagnetic *A* phase below $T_C^M$ and the value of $T_C^M$ is not experimentally observable anymore. For clarity, we have extrapolated the $T_C^A$ and $T_C^M$ values and indicated them in the inset of fig. 2 by open pentacles (☆).

According to the results above, it is clearly observed that the shift of martensitic-transformation temperature in this Mn-poor $Mn_{1-x}CoGe$ system strongly correlates with the deficiency level ($x$) of Mn composition, while with a weak Mn deficiency-level dependence of Curie temperatures of austenite and martensite. In the MnCoGe crystal structure, the Co and Ge atoms form three-dimensional chain networks via covalent bonds owing to the *sp* and *sd* orbital hybridization [11,24]. The existence of Co vacancies in the Co-Ge-Co chains inevitably destroys a part of the chemical bonds, which may result in destabilization of the *M* phase at lower temperature and in weakening of the elastic modulus of the *A* phase [25,26]. The phase transformation thus needs a smaller thermodynamic driving force to overcome the energy barrier of *M* nucleation [27,28]. As shown in table 1, $\Delta T$ of the martensitic transformation indeed exhibits a decreasing trend, from 22 to 12 K, as a function of increasing vacancy content. In conclusion, the Co vacancies induce regular decrease of $T_m$, the same as reported for Co-poor MnCoGe alloys [10,15,16,18], and of hysteresis $\Delta T$ of transformation in $Mn_{1-x}CoGe$ alloys.

In the Mn-poor $Mn_{1-x}CoGe$ system, a conspicuous difference with respect to the Co-poor $MnCo_{1-x}Ge$ system is that the values of the Curie temperatures of martensite and austenite depend only weakly on the level of the Mn deficiency ($x$). If the Mn deficiency changes, the valence-electron concentration (*e/a*) determined as the concentration-weighted sum of *s*, *d*, and *p* electrons, will also change in the system. We propose *e/a* as a tentative parameter to examine the change of Curie temperatures, as shown in fig. 3. It can be seen that the Curie temperatures of both phases remain almost unchanged, whereas $T_m$, exhibits a strong dependence on *e/a* in the present *e/a* range. The weak *e/a* dependence of the Curie temperature suggests there is only weak change of the ferromagnetic exchange in both the martensite and the austenite phases which, according to the spin-fluctuation theory [29], may be ascribed to a small difference of the electronic structures. The similar features have been reported in Heusler-type NiMn-based alloys with phase transitions [30,31]. From the inset of fig. 3, one can see that, in $Mn_{1-x}CoGe$ alloys, the *e/a* change with decreasing Mn content is very small, whereas, in sharp contrast, the *e/a* change with the same decrease of the Co content is much larger in $MnCo_{1-x}Ge$ alloys. This may account for the stable values of $T_C^M$ and $T_C^A$ in the $Mn_{1-x}CoGe$ system and deserves further study.

In fig. 4, we present the thermomagnetization of the $Mn_{1-x}CoGe$ alloys of the temperature window in a high field of 5 T. A large $\Delta M$ is associated with the martensitic transformation with values as listed in table 1. In contrast to the other alloys, the alloys in the window exhibit quite abrupt increments of the magnetization due to the transformation from the PM *A* phase to the FM *M* phase with $\Delta M$ values up to about 60 $Am^2/kg$. This value is about twice as large as the value of about 32 $Am^2/kg$ of Co-poor MnCoGe at the FM/FM-type martensitic transformation [15]. For the alloy with $x$ = 0.045, along with the transformation temperature decreasing out of the window, i.e. $T_m < T_C^A$, the transformation occurs from FM austenite to FM martensite and the $\Delta M$ reduces to only 34 $Am^2/kg$ . For the alloys in the window, it can be seen their $T_m$ shifts toward to the high-temperature region in high field relative to that in low field shown in fig. 2, due to the higher magnetization of the FM *M* phase.

Concerning the large value of $\Delta M$ at the transformation in the window, we present in fig. 5 isothermal $M$ ($H$) curves for the alloy with $x$ = 0.035, measured upon cooling in fields up to 13 T. From 300 down to 285 K, a metamagnetic phase transition with hysteresis indicates that a magnetic-field-induced martensitic transformation occurs in this temperature region. The large $\Delta M$ gives rise to a large difference in Zeeman energy between the PM $A$ and the FM $M$ phases which drives the transition. At 5 K, the curve shows typical ferromagnetic magnetization behavior of the $M$ phase with a high saturation magnetization of 109 $Am^2/kg$.

The magnetocaloric effect for the alloy with $x$=0.035 has been derived from magnetization curves measured in fields up to 5 T. Figure 6 shows that, around $T_m$, the magnetic entropy $\Delta S_m$ reaches giant and negative values. Maximum values of -10 J/kg·K for ΔB = 0 - 2 T and -26 J/kg·K for ΔB = 0 - 5 T are observed, which are comparable with values for other materials in a similar temperature range, such as $Gd_5Si_2Ge_2$ (-14 J/kg·K for ΔB = 0 - 2 T) [32], $Ni_{50}Mn_{33.13}In_{13.90}$ (-28.6 J/kg·K for ΔB = 0 - 5 T) [33] and $MnCoGe_{0.95}Sn_{0.05}$ alloys (-4.5 J/kg·K for ΔB = 0 - 1 T) [21]. It is important that, in this PM/FM-type martensitic transformations, the exothermic/endothermic effect corresponding to the magnetic-entropy change of these alloys has the same sign as that of the martensitic structural transformation, which enhances the caloric effect.

In summary, it is shown that PM/FM-type martensitic transformations with substantial magnetization changes can be realized in Mn-poor $Mn_{1-x}CoGe$ alloys in the range of $0.010 < x < 0.045$ with similar associated Co-vacancy contents. By varying $x$, these martensitic transformations are tunable in a temperature window between $T_C^A$ and $T_C^M$ . Also, a field-induced martensitic transformation and large magnetocaloric effect are reported. The martensitic-transformation temperature of the $Mn_{1-x}CoGe$ alloys strongly depends on the Mn content whereas the Curie temperatures of the martensite and austenite remain almost unchanged. The reported vacancy-tuning via Mn deficiency may also be of interest in case of other materials in the MM'X family.

***

This work is supported by the National Natural Science Foundation of China in Grant No. 50531010 and 50771103.

# FIGURE CAPTIONS

Fig. 1: (Color online) Power XRD patterns of the $Mn_{1-x}CoGe$ alloy with $x$=0.035 at 320 and 220 K upon cooling. The Miller indices $hkl_h$ and $hkl_o$ denote the hexagonal and orthorhombic structure, respectively. The unit cells of both structures are also shown.

Fig. 2: (Color online) Temperature dependence of the magnetization of $Mn_{1-x}CoGe$ ($x$= 0.010, 0.030, 0.035 and 0.045) alloys in a field of 0.01 T. $M_s(T_m)$ and $M_f$ denote the starting and finishing temperatures of the martensitic transformation, $A_s$ and $A_f$ the starting and finishing temperatures of the reverse transformation, while the $T_C^A$ and $T_C^M$ denote Curie temperatures of $A$ and $M$ phases, respectively. The inset shows the phase diagram for the $Mn_{1-x}CoGe$ system. The solid pentacles denote the measured $T_C$ values and the open pentacles extrapolated $T_C$ values.

Fig. 3: (Color online) Valence-electron concentration ($e/a$) dependence of the martensitic transition temperature ($T_m$), Curie temperatures of austenite and martensite ($T_C^A$ and $T_C^M$) of $Mn_{1-x}CoGe$ alloys. The inset shows the vacancy-content ($x$) dependence of $e/a$. The value of $e/a$ has been determined as the concentration-weighted sum of $s$, $d$, and $p$ electrons.

Fig. 4: (Color online) Temperature dependence of the magnetization of $Mn_{1-x}CoGe$ ($x$=0.010, 0.030, 0.035 and 0.045) alloys in a field of 5 T.

Fig. 5: (Color online) Isothermal magnetization curves of the alloy with $x$=0.035 at various temperatures across the martensitic transformation.

Fig. 6: (Color online) Isothermal magnetic-entropy changes derived from magnetic isotherms of the alloy with $x$= 0.035, measured in fields up to 5 T.

Table 1: Composition dependence of $T_m$, $T_C$, $\Delta M$ and $\Delta T$ for $Mn_{1-x}CoGe$ alloys.

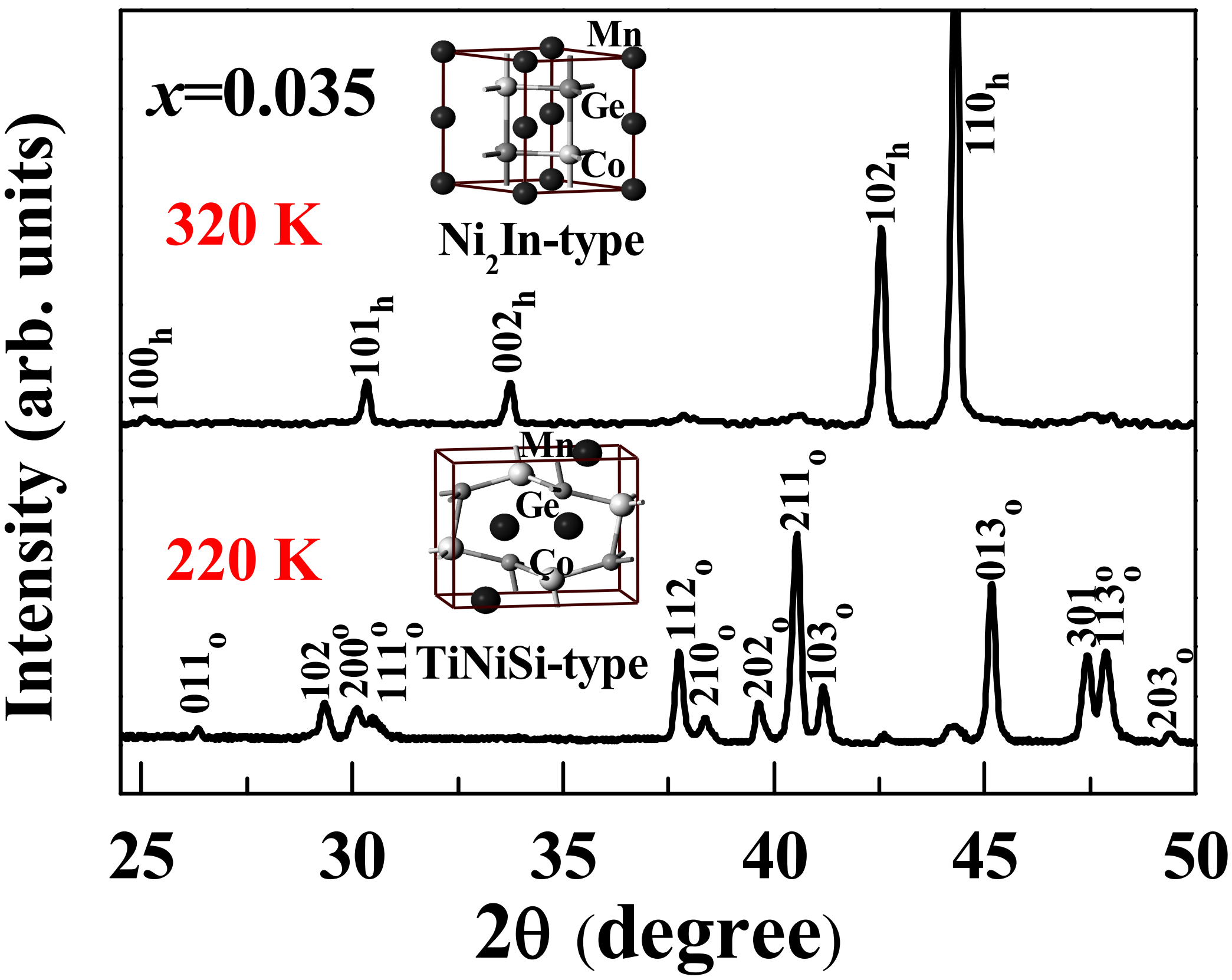

Fig. 1

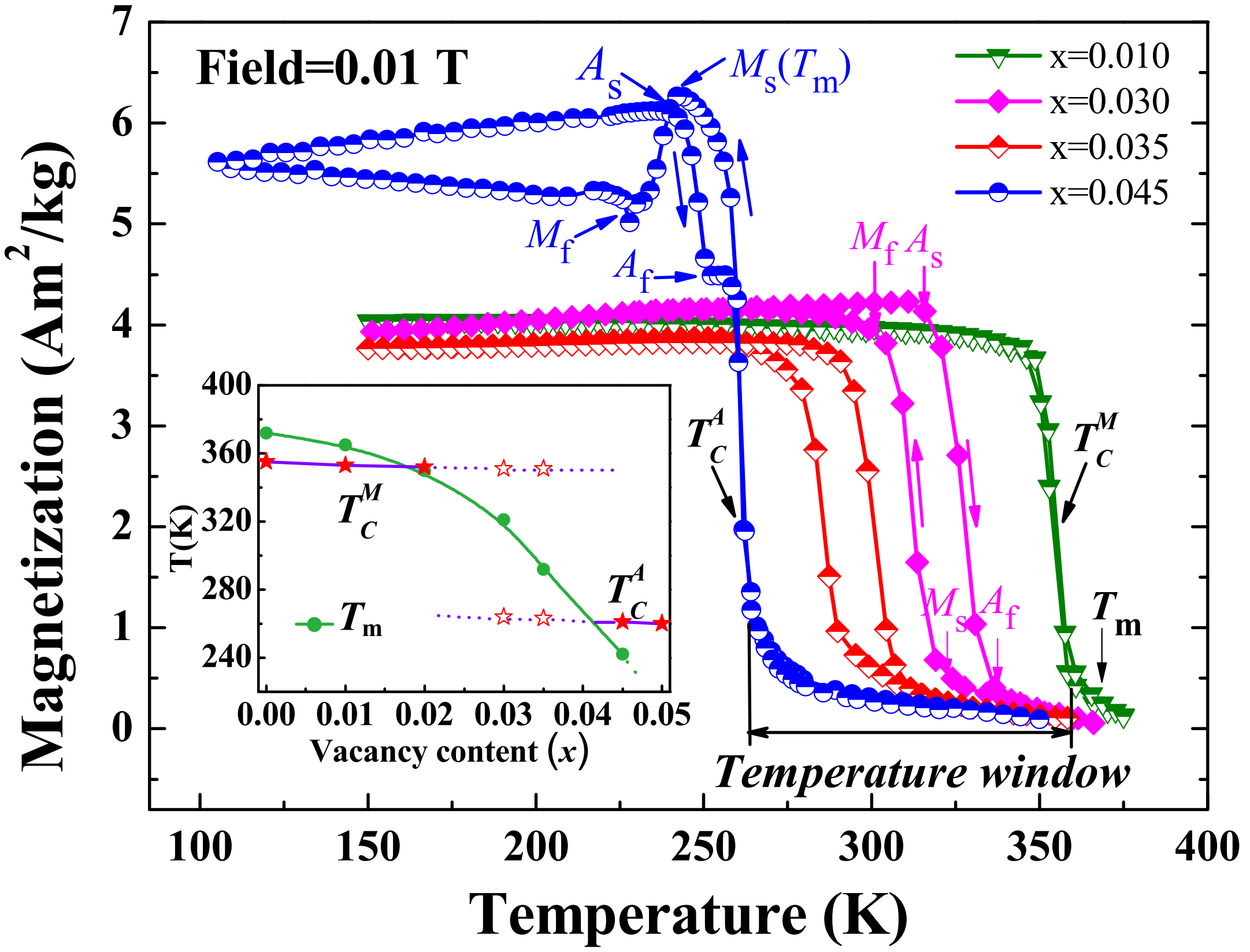

Fig. 2

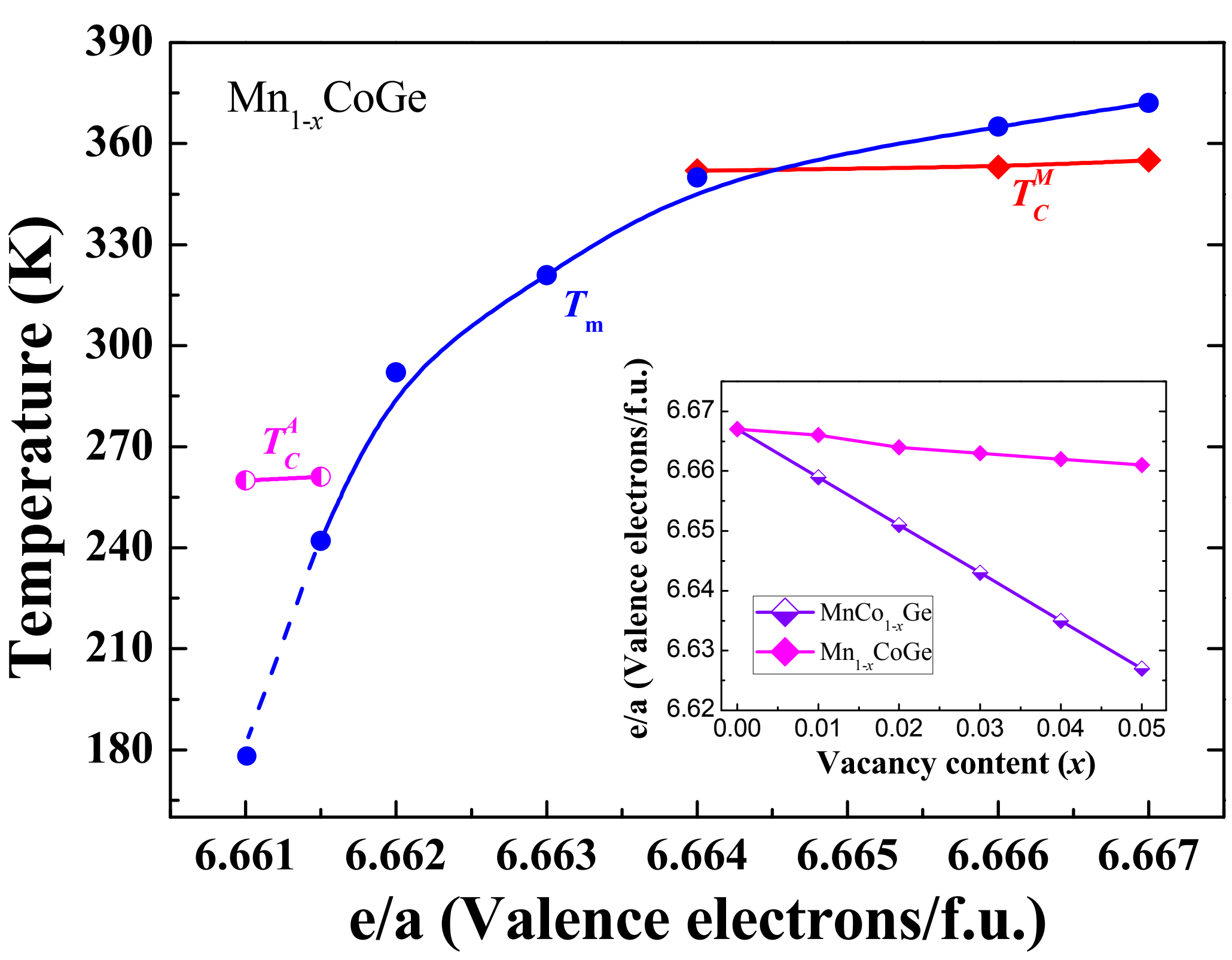

Fig. 3

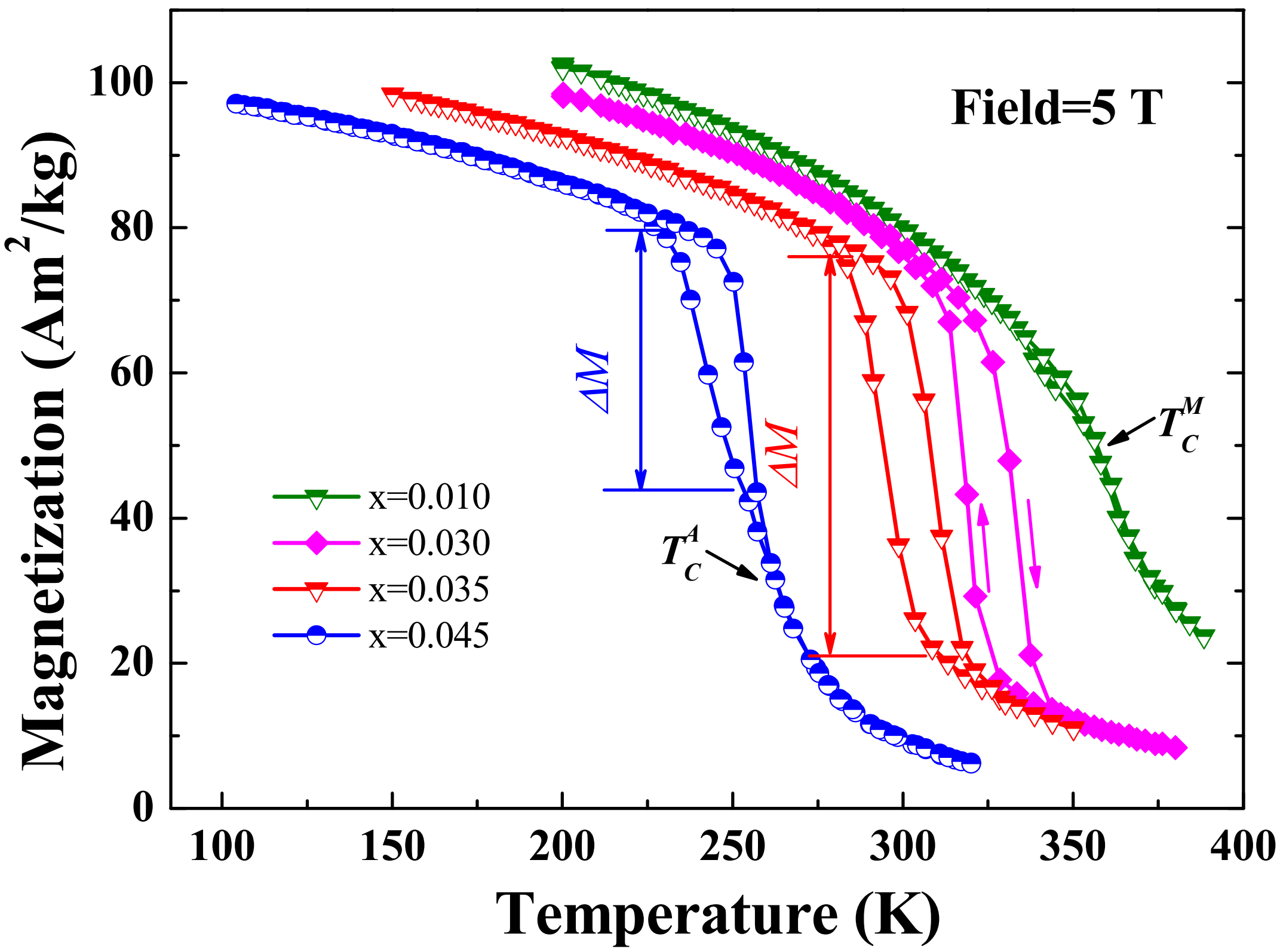

Fig. 4

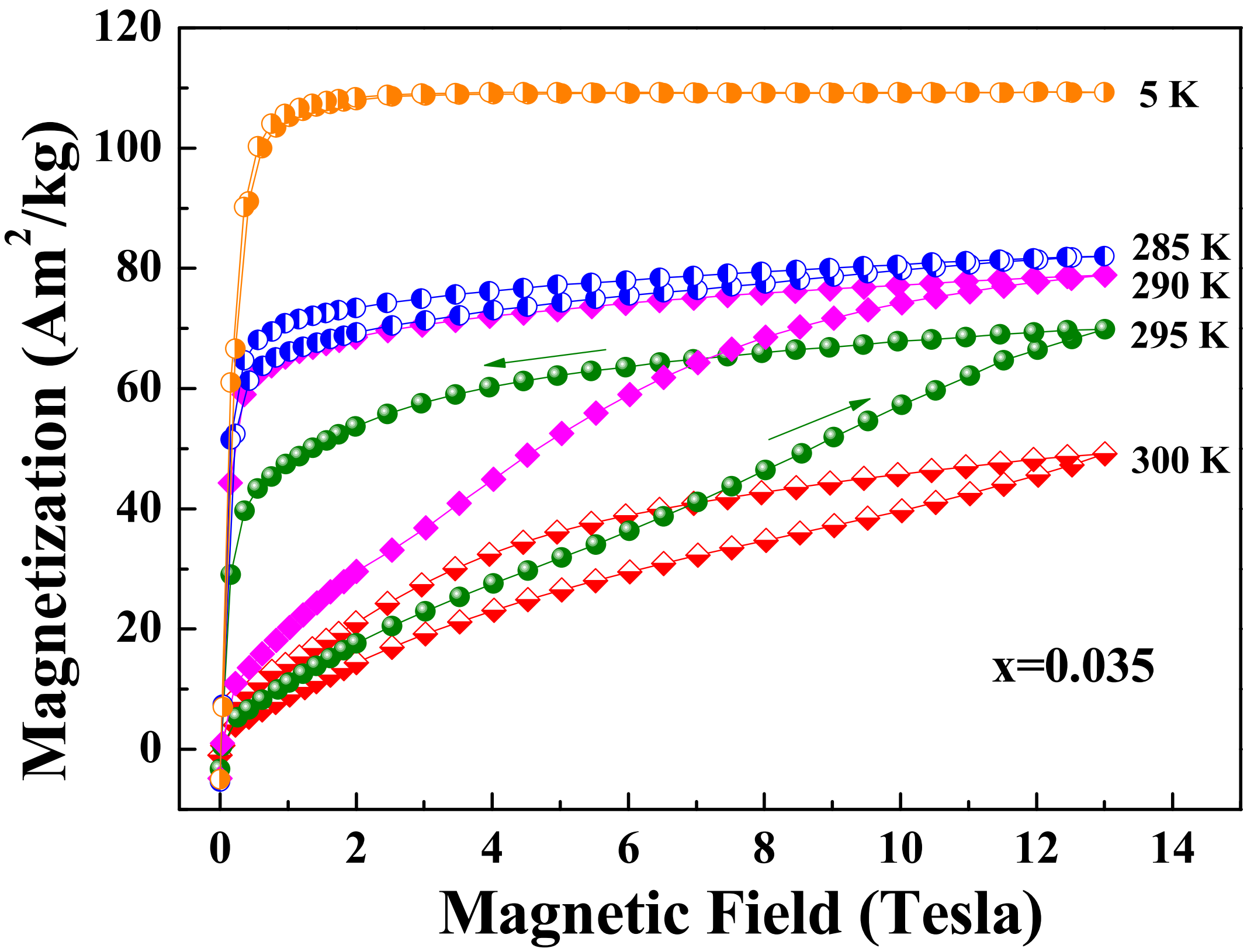

Fig. 5

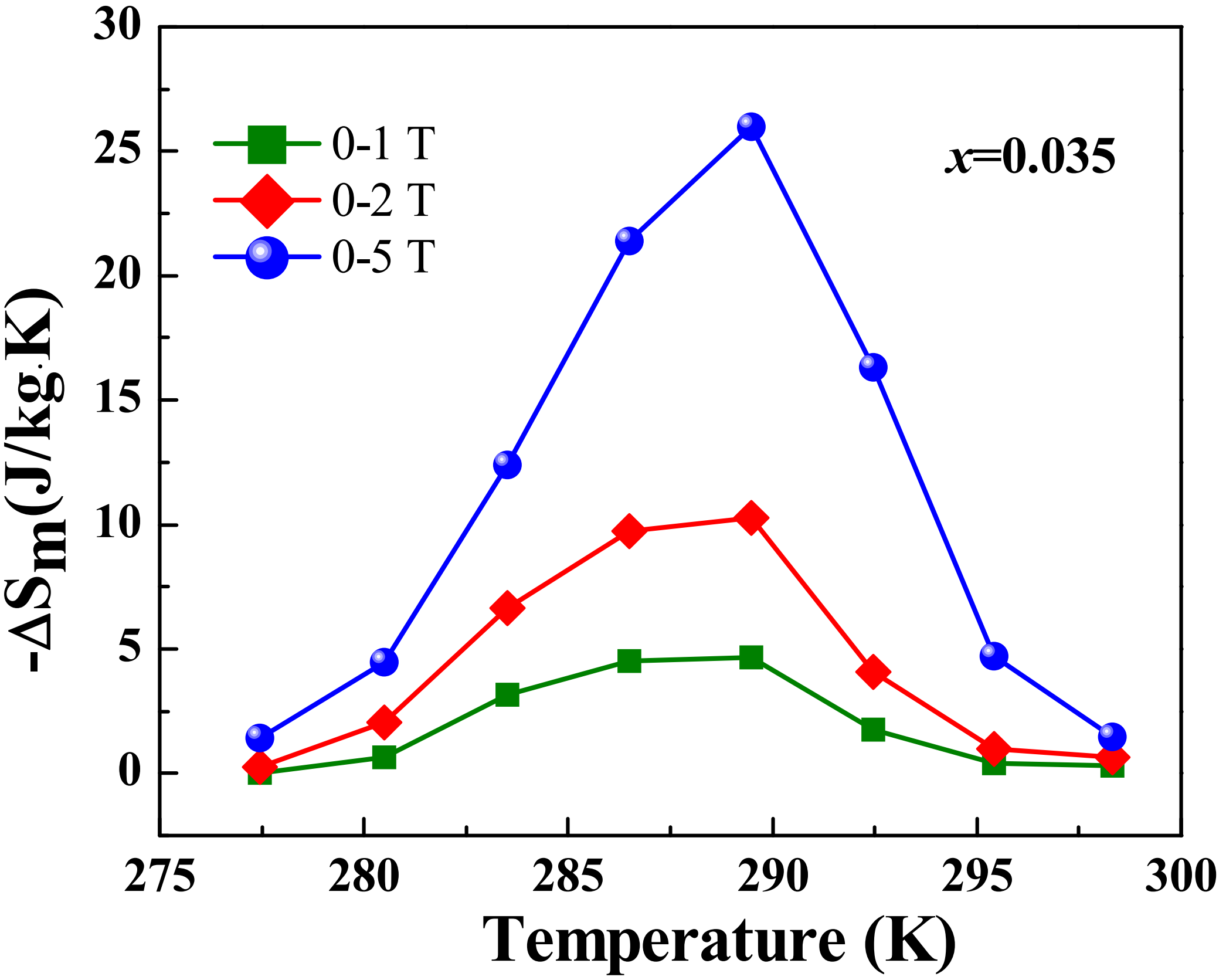

Fig. 6

| $x$ | $T_m$(K) | $T_C$(K) | $\Delta M$(Am$^2$/kg) | $\Delta T$(K) |
|---|---|---|---|---|
| **0.000**[a] | 372 | 355 | - | 29 |
| **0.010**[a] | 365 | 353 | - | 22 |
| **0.020**[b] | 350 | 352 | 60 | 19 |
| **0.030**[c] | 321 | - | 58 | 16 |
| **0.035**[c] | 292 | - | 59 | 15 |
| **0.045** | 242 | 261 | 34 | 12 |
| **0.050**[d] | - | 260 | - | - |

a. The $T_m$ and $\Delta T$ are determined using DTA as the transformation occurs at the temperature range where the system is in the paramagnetic state. The $\Delta M$ are absent due to the same reason.

b. The data of this alloy are determined using DTA and magnetic measurements as its transformation is situated at the $T_C$ of martensite.

c. The $T_C$ are immeasurable due to the PM/FM-type martensitic transformation behaviors of these two alloys.

d. The $T_m$, $\Delta M$ and $\Delta T$ are absent as the martensitic transformation behavior of this alloy vanishes after annealing.

# Table 1